\documentclass[submission,copyright,creativecommons]{eptcs}
\providecommand{\event}{GandAlf 2017}
\usepackage{breakurl}
\usepackage{underscore}
\usepackage{graphicx}
\usepackage{caption}
\usepackage{subcaption}

\usepackage{amsmath}
\usepackage{amssymb}

\usepackage{framed}
\usepackage{float}
\usepackage{color}
\usepackage{verbatim}

\usepackage{url}

\usepackage{algorithm}
\usepackage[noend]{algpseudocode}

\newtheorem{theorem}{Theorem}

\newtheorem{definition}[theorem]{Definition}

\title{ParaPlan: A Tool for Parallel Reachability Analysis of Planar Polygonal Differential Inclusion Systems}

\author{
    Andrei Sandler
    \institute{School of Computer Science\\University of Hertfordshire\\United Kingdom}
    \email{a.sandler@herts.ac.uk}
\and
    Olga Tveretina
    \institute{School of Computer Science\\University of Hertfordshire\\United Kingdom}
    \email{o.tveretina@herts.ac.uk}
}

\def\titlerunning{A Tool for Parallel Reachability Analysis of Planar Polygonal Differential Inclusion Systems}
\def\authorrunning{A. Sandler \& O. Tveretina}

\graphicspath{ {images/} }

    {\framed\verbatim}%
    {\endverbatim\endframed}

\begin{document}
\maketitle

\begin{abstract}

\textbf{Abstract.}
    We present the ParaPlan tool which provides the reachability analysis of planar hybrid systems defined by 
    differential inclusions (SPDI). It uses the parallelized and optimized version of the algorithm underlying the SPeeDI tool \cite{APSY02}. The performance comparison demonstrates the speed-up of up to 83 times with respect to the sequential implementation on various benchmarks.
    Some of the benchmarks we used are randomly generated with the novel approach based on the partitioning of the plane with Voronoi diagrams.

\end{abstract}

\section{Introduction}
    A hybrid system is a dynamic system that exhibits both continuous and discrete behaviour. Examples of such systems come from robotics, avionics, air traffic management and automated highway management. Most of the hybrid systems are safety critical and errors can have serious consequences. Formally, verifying safety properties of hybrid systems consists of building a set of reachable states and checking if this set intersects with a set of unsafe states. Therefore one of the most fundamental problems in the analysis of hybrid systems is the reachability problem.

    The reachability problem is only decidable for special classes of hybrid systems \cite{HKPV98}. Currently, a number of tools for analysing the reachability problem are available, including dReach \cite{KGCC15}, Flow$^*$ \cite{CAS13}, KeYmaera \cite{PQ08} and HSolver \cite{RS05}.
    

    The focus in developing tools for the reachability analysis is now mainly on improving the performance of sequential algorithms, because the sequential algorithms do not always provide the required computational efficiency. Approaches for parallelisation are still uncommon and the benefits of parallel execution on multi-core platforms are not well understood \cite{SAMFSK15}. The main motivation for our work is to understand further computational benefits of parallelization of the reachability analysis.

    In this paper we consider a decidable class of hybrid systems, called Planar Polygonal Differential Inclusions (SPDIs), which naturally arises from the analysis of hybrid systems with two continuous variables. SPDIs are defined by giving a finite partitioning of the plane into convex polygonal sets, together with a differential inclusion associated with each region $P$ and defined by a couple of vectors $l_{P}$ and $r_{P}$. An algorithm for solving the reachability problem for SPDIs has been introduced in \cite{ASY01}. It abstracts trajectory segments into so-called signatures (sequences of edges and simple cycles) and then even further into types of signatures (signatures which do not take into account the number of times each simple cycle is iterated).

    In \cite{APSY02,speedi06} the authors present the SPeeDI toolkit which is a collection of utilities to manipulate and reason automatically about SPDIs. The tool is implemented in Haskell and also provides trace generation on top of the reachability analysis, but it was never benchmarked or optimized.

    The decidability result for SPDIs has also been extended to generalized SPDIs in \cite{PS08}. Those are SPDIs not satisfying the goodness assumption (the dynamics of a region of the SPDI do not allow a trajectory to traverse an edge in opposite directions).
    
    There is a generalized version of the SPeeDI tool, namely GSPeeDI \cite{hansen2009gspeedi}, written in Python. It computes all simple cycles using the algorithm of Tarjan \cite{T73}. Obviously, the number of simple cycles is the bottleneck determining when the problem becomes infeasible.

    The choice of SPDIs has been triggered by two factors: on the one hand this class of hybrid systems is decidable and on the other hand it is powerful enough to exhibit relatively complex behaviour. Moreover, SPDIs cannot be straightforwardly verified by the existing tools due to non-determinism expressed by differential inclusions.

    {\it Contribution.} Our contribution is twofold. First, we present the ParaPlan tool for PARAllel analysis of PLANar differential inclusion systems which implements the optimized and parallelized version of the sequential algorithm underlying the SPDI tool. Second, we describe the novel approach for random generation of benchmarks using Voronoi diagrams. ParaPlan is available online at \cite{ST17}. It has been tested on a series of benchmarks, including those from \cite{speedi06} and random benchmarks generated using our approach. Absolute testing time and relative speed-up against original algorithm is measured.

    {\it Related work.} To the best of our knowledge, the results on parallelization of the reachability problem for hybrid systems were reported only in \cite{PS2006} and \cite{GDBBGR16}.

    In the earlier paper \cite{PS2006}, although purely theoretical, the authors introduce a compositional algorithm for splitting the reachability task into several independent tasks in the strongly connected regions of an SPDI. We decided not to implement this algorithm because the only case it can speed up calculations is when multiple tasks are solved consequently on the same SPDI, and pre-calculations take as much time as it is needed to solve one reachability task with the original algorithm.

    In \cite{GDBBGR16} two parallel state-space-exploration algorithms have been proposed for the reachability analysis of general hybrid systems, which are implemented in the XSpeed model checker. The first algorithm uses the parallel, breadth-first-search algorithm of the SPIN model checker. The second algorithm improves load balancing. Their approach is to parallelize BFS algorithm and divide calculations inside a discrete state into 'atomic' tasks for better load balancing between threads. In case of the ParaPlan tool the state space is divided into atomic tasks (dynamic flow on the region's edges) by design.

    Although there are many tools available for hybrid systems analysis, their comparative evaluation is problematic as they do not support the same model classes. Moreover, it is impossible to use those tools on SPDI directly, because it is not allowed to use differential inclusions inside a discrete state.

    {\it Outline.} In Section \ref{sec:SPDI} we formally describe the class of two-dimensional non-deterministic hybrid systems studied in this paper, namely SPDIs. In Section \ref{sec:SA} we recall the original approach for computing reachable states for SPDIs introduced in \cite{ASY07,APSY08} and in Section \ref{sec:optimisation} we present our optimisation of the algorithm and its parallel version. In Section \ref{sec:benchmarks} we describe the novel approach for generating random benchmarks. Performance evaluation can be found in Section \ref{sec:tests}. Section \ref{sec:conclusions} contains concluding remarks.

\section{Polygonal Differential Inclusions}\label{sec:SPDI}

    The notion of an SPDI is a generalization of Piecewise-constant Derivative Systems (PCD) studied in \cite{AMP95}.   The new characteristic of SPDIs with respect to PCDs is non-determinism. Informally, an SPDI consists of a partition of a plane subset into convex polygonal regions, together with a differential inclusion associated with each region \cite{ASY01}. That is, the class of SPDI systems can be represented as non-deterministic linear hybrid automata with continuous trajectories which derivative in every point inside any convex region is bounded by a given angle.

    Now we will define an SPDI formally, and we will use $\angle_{r}^{l}$ to denote the angle defined by two non-zero vectors $\vec{l}$ and $\vec{r}$.
    
    \begin{definition}
        We define an \textit{SPDI} as a hybrid automaton $H = (Q, D, X, f, E, Init, G, R)$, 
        where

        \begin{itemize}
            \item $Q = \{q_{1}, q_{2}\, \dots, q_n \}$ with $n>0$ is a finite set of discrete states (regions);
            \item $X = \mathbb{R}^{2}$ is a set of continuous states;
            \item $f(\cdot,\cdot): Q\times X \rightarrow X$ is a 
            vector field bounded in every region $q_{i}$ by $\angle_{r_i}^{l_i}$;
            \item A domain function $D(\cdot):Q\rightarrow P(X)$ defines regions as a set of convex (possibly infinite) polygons, forming a convex polygon partitioning of $\mathbb{R}^{2}$;
            \item $Init$ is a set of edge intervals;
            \item $E$ is a set of edges between regions, formed by all polygon boundaries;
            \item A guard condition $G(\cdot):E\rightarrow P(X)$ is a linear guard condition, defined by the edges of the partitioning;
            \item A reset map $R(\cdot,\cdot): E\times X\rightarrow {{P}}(X)$ is an identity function.
        \end{itemize}
    \end{definition}

\begin{figure}[h]
        \centering
        \includegraphics[scale=0.27]{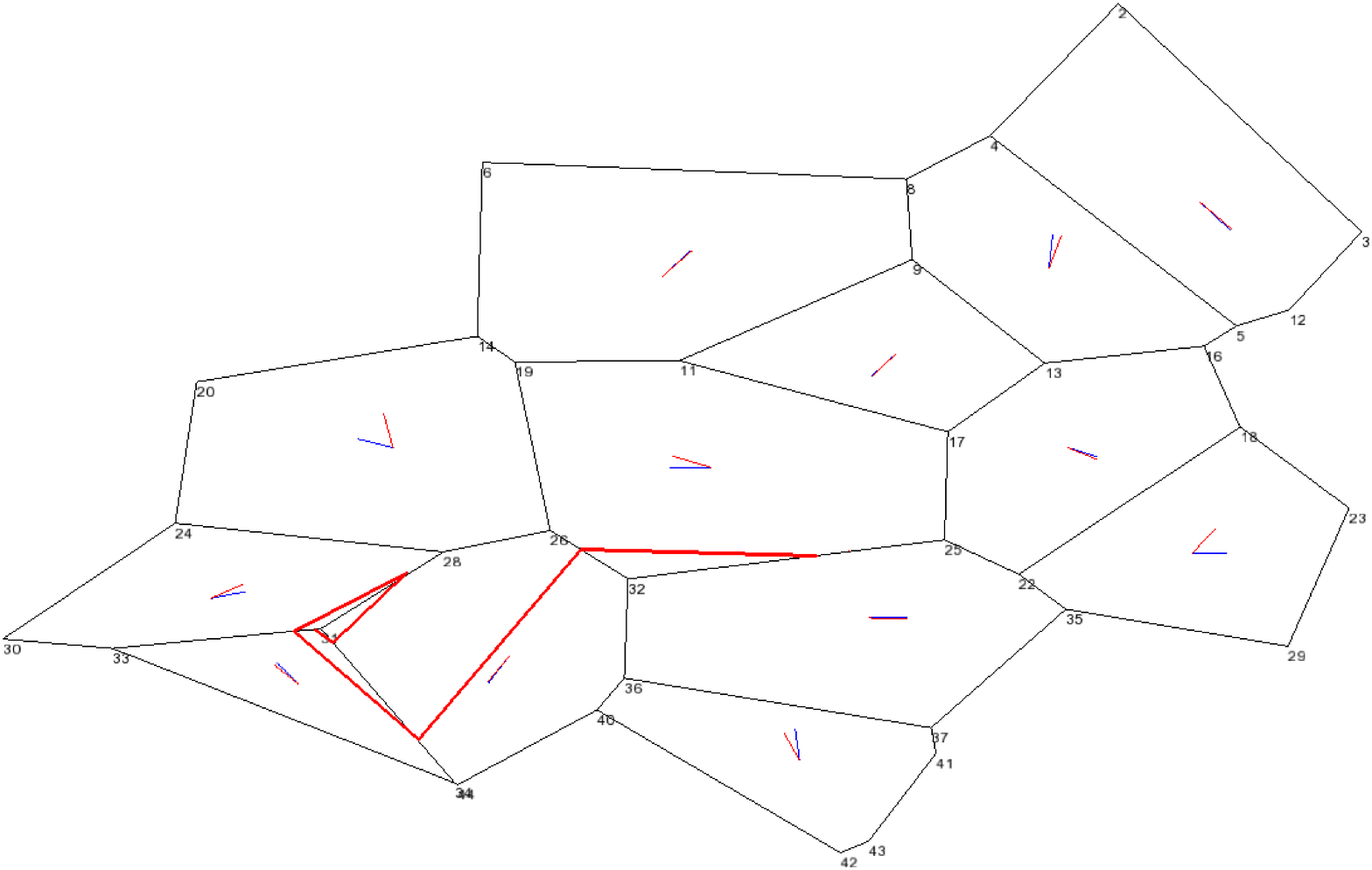}
            \caption{The SPDI with 13 regions and a trajectory segment}
        \label{fig:spdi-example}
    \end{figure}

    Figure \ref{fig:spdi-example} illustrates the SPDI randomly generated using the method described in Section \ref{random-spdi-generation} and a trajectory segment.

    The edge-to-edge reachability problem for SPDI has been proved to be decidable (\cite{ASY07}) and could be stated as follows:

    \textit{Given two edges $e_0$ and $e_f$, does there exist $x_0\in e_0$ and $x_f\in e_f$ such that there is a trajectory segment starting at $x_0$ and ending at $x_f$?}

    This task could be interpreted as following: if the whole model represents the dynamics of a real-world system, a trajectory represents one possible evolution of the system, and there are some unsafe states, then the existence of a trajectory starting in an initial set of states and ending in an unsafe state proves such system to be unsafe to use.

\section{Sequential Algorithm}\label{sec:SA}

    In this section we recall the original approach for computing reachable states, which is introduced in \cite{ASY07,APSY08} and based on the characterization of the qualitative behaviours of trajectories.

    In general, there are infinitely many trajectories from the starting set $S$ to the final set $F$, but they all are determined by the angles associated with the regions $q_{i}$. Therefore, for edges $e, e' \in q_i$ and an interval $(s, s') \subseteq e$, there is an interval $(f, f') \subseteq e'$, such that every trajectory starting in $(s, s')$ will intersect with $(f, f')$. Hence, we can calculate the reachable states just by examining the edge intervals that the trajectories traverse. If a trajectory crosses some intervals successively on the edges $e_{1}, e_{2}, \dots, e_{n}$, then the sequence $\sigma = (e_{1}, e_{2}, \dots, e_{n})$ is called the \textit{edge signature}.

    For computing the successive interval images, it is convenient to introduce a one-dimensional coordinate system on each edge $e$, with zero (0) denoting one chosen vertex $v_{0}$ of $e$ and one (1) denoting the other vertex $v_{1}$. Now each point between the vertices of each edge has the coordinate $v_{\lambda} = \lambda v_{0} + (1 - \lambda) v_{1}$ with $0<\lambda <1$. Then, a series of successor functions on edges of the SPDI is defined.

    \begin{itemize}
        \item The  successor $Succ_{\textbf{c}}(x)$ of a point $x$ on an edge $e$  under a dynamics, defined by a single vector $\textbf{c}$ is an image $x'$ on the edge $e'$ of the same region, where the point $x$ will be projected along $\textbf{c}$.

        \item The successor $Succ_{(\textbf{c}_{1}, \textbf{c}_{2})}(x_{1}, x_{2})$ of an interval $(x_{1}, x_{2})$ on edge $e$ in region $r$ with dynamics, defined by $\angle_{c_2}^{c_1}$  is an interval $(x_{1}', x_{2}')$ on $e'$, where
           $$x_{1}' = min(1, Succ_{\textbf{c}_{1}}(x_{1}), Succ_{\textbf{c}_{2}}(x_{1}))$$
           $$x_{2}' = max(0, Succ_{\textbf{c}_{1}}(x_{2}), Succ_{\textbf{c}_{2}}(x_{2}))$$
        If $x_{1}' > x_{2}'$, then the successor is the empty set. In other words, the successor of an interval is the interval  reachable under the region's dynamics.

        \item The successor $Succ_{\sigma}(x_{e_{1}, 1}, x_{e_{1}, 2})$ of an interval $(x_{e_{1}, 1}, x_{e_{1}, 2})$ on edge $e_{1}$ along the edge signature $\sigma = (e_{1}, e_{2}, \dots, e_{n})$, is a result of applying $Succ_{(\textbf{c}_{e_{i}, 1}, \textbf{c}_{e_{i}, 2})}(x_{e_{i}, 1}, x_{e_{i}, 2})$ consequently to $e_{1}, e_{2}, \dots, e_{n-1}$.
        
        Roughly speaking, the successor of an interval along $\sigma$ is the set of points on $e_n$  reachable from the points on $e_1$ through $e_{2}, e_{3}, \dots, e_{n-1}$.
    \end{itemize}

    The edge signature of a trajectory can possibly contain simple cycles, but nested cycles are not permitted. In general, an edge signature has the following form:

    \begin{equation*} \label{eq:sig}
        r_{1}s_{1}^{k_{1}}r_{2}s_{2}^{k_{2}} \dots r_{n}s_{n}^{k_{n}}r_{n+1}
    \end{equation*}
    where $r_{i}$ denotes the path between cycles, and $s_{i}^{k_{i}}$ denotes the cycle $s_{i}$ repeated $k_{i}$ times.

    Still, an SPDI can have infinitely many edge signatures, because in some trajectories there are cycles $s_{i}$ that could be repeated any $k_{i}$ times. In order to compactify all edge signatures into a finite set, it is generalized again using signature types.

    The \textit{signature type} of an edge signature is the following sequence:
    \[
        r_{1}s_{1}r_{2}s_{2} \dots r_{n}s_{n}r_{n+1}
    \]

    The following theorem defines a set of edge signatures, which is only needed to be examined for finding the trajectory from $S$ to $F$.

    \begin{theorem}[Asarin, Schneider, Yovine, \cite{ASY07}]\label{theorem:signature}
        Only those signature types having disjoint paths $r_{i}$ and unique (as sets of edges) cycles $s_{i}$, could correspond to a trajectory starting in initial set $S$ and ending in final set $F$. It is easy to see that there are only finite number of such signature types on any given SPDI.
    \end{theorem}

    Having fixed a signature type and starting intervals on the first edge of the signature type, one can algorithmically calculate the successor function along the edges of the signature type and check if the final set could be reached. Cycles are treated in a special way described in \cite{ASY07}. Following from the existence of the deciding algorithm, the reachability problem on SPDI is decidable.

\section{Optimization of Sequential Algorithm}\label{sec:optimisation}

\begin{figure}[h]
        \centering
        \includegraphics[scale=0.4]{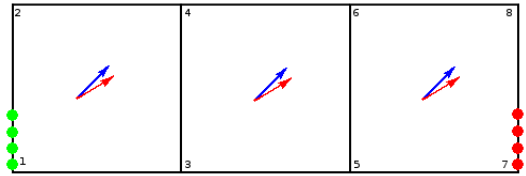}
        \captionsetup{justification=centering}
        \caption{The red interval is not reachable from the green interval,\\ but the corresponding edge signature is feasible}
        \label{fig:not-reachable}
    \end{figure}

    The algorithm underlying the SPeeDI tool constructs all feasible types of signatures. If even the time for analysing each signature type is not significant, it computes also the signature types which cannot be realised in any trajectory. The example of such behaviour is given in Figure \ref{fig:not-reachable}.

    In our approach we explore feasible signatures via DFS and simultaneously compute reachable states. This way we look up less or equal number of signatures, because every explored signature would have a trajectory realization (because of reachability). During the edge exploration we check the conditions of Theorem \ref{theorem:signature} and do not take into consideration those edges that form nested or non-unique cycles.

    For effective analysis of cycles in signatures we use the technique described in \cite{ASY07}. There are only five different types of cycle behaviour, and the exit sets of points in four cases out of five could be effectively calculated without iterating the cycle. The last cycle type requires iteration, but it is proved to be finite.

    The main algorithm for signature types discovering is listed below. To solve the whole reachability task, one simply need to explore all the signature types from every starting edge:

\begin{algorithm}
\caption{Solving the reachability problem}

\begin{algorithmic}
\Function{SolveReachabilityTask}{spdi, reachTask}
  \For{startEdge \textbf{in} reachTask.StartEdges}
    \If {DFSSignaturesExploration(startEdge, spdi, reachTask)}
      \Return SUCCESS
    \EndIf
  \EndFor

  \State \Return FAILURE
\EndFunction
\end{algorithmic}
\end{algorithm}

    In the main DFS function we use the fact (Theorem \ref{theorem:signature}) that all already visited edges must belong to the last discovered path $r_{i}$, otherwise there will be a nested cycle.

\newpage

\begin{algorithm}
\caption{Main depth-first search function for exploring the signatures of trajectories}
\begin{algorithmic}
\Function{SignaturesExploration}{currentEdge, borders, spdi, reachTask}
  \If{final state on currentEdge is reached}
    \Return SUCCESS
  \EndIf

  \If{currentEdge is visited}
    \State iterate back to restore cycle and current path $r_{i}$

    \If{cycle is visited \textbf{OR} ends not in current path $r_{i}$}
      \State \Return FAILURE \Comment{only simple cycles allowed}
    \EndIf

    \State mark new cycle as visited
    \State reachable $\gets$ \Call{TestCycleAndGetFinalImages}{cycle, borders, spdi, reachTask}

    \If{final state is reached in reachableStates}
      \Return SUCCESS
    \EndIf

    \For{image \textbf{in} reachable}
      \If{image is not valid}
        \Return FAILURE
      \EndIf
      
      \For{all possible nextEdge, connected to currentEdge}
        \State nextImage $\gets$ \Call{SuccInt}{image, currentEdge, nextEdge}
        \If{nextImage is valid}
          \State \Return \Call{SignaturesExploration}{nextEdge, nextImage, spdi, reachTask}
        \EndIf
      \EndFor
    \EndFor
  \Else
    \State mark currentEdge visited
    \State add currentEdge to current path $r_{i}$

    \For{all possible nextEdge, connected to currentEdge}
      \State nextImage $\gets$ \Call{SuccInt}{borders, currentEdge, nextEdge}
      \If{nextImage is valid}
        \State \Return \Call{SignaturesExploration}{nextEdge, nextImage, spdi, reachTask}
      \EndIf
    \EndFor
  \EndIf

  \Return FAILURE
\EndFunction
\end{algorithmic}
\end{algorithm}

    Auxiliary function \texttt{TestCycleAndGetFinalImages} determines the type of cycle (one of \texttt{\{STAY, EXIT-LEFT, EXIT-RIGHT, EXIT-BOTH, DIE\}}, see \cite{ASY07}), and effectively calculates the set of intervals of the first cycle edge which will be visited during cycle iteration.

\subsection{Parallelization}\label{sec:parallelization}

    We further improve the sequential algorithm by parallelizing the DFS-like exploration of signature types. We do it by assigning sub-trees of DFS to different threads and loading them again by new sub-trees when they are finished with previous tasks. When the algorithm iterates over all possible next edges in the signature type, except the last, it will check if there is a free thread which could be loaded with the DFS sub-tree starting with this edge. Mutexes are used to eliminate the possible race condition.

    During the computation, some sub-tasks may finish earlier than others. In this case the remaining sub-tasks will be divided to load the free threads again, so the CPU utilisation will be full all the time.

    Data is partly shared between threads (such as SPDI representation and reachability task), but partly it is copied when creating new threads, because it cannot be stored globally. For example, visited edges and cycles are different at almost all times in each two different threads threads, therefore storing it in one data structure will give no gain in memory or time consumption.
    
\begin{algorithm}
\caption{Sequential algorithm parallelization}
\begin{algorithmic}
\State \textbf{global} FreeThreads = NumberOfThreads

\Function{SolveReachTask}{spdi, reachTask}
    \For{$e$ \textbf{in} reachTask.StartEdgeParts}
        \If{FreeThreads $>$ 0 \textbf{AND} $e$ is not the last}
            \State FreeThreads $\gets$ FreeThreads - 1
            \State \Call {CreateThread}{SignaturesExploration, $e$.edge, $e$.borders, spdi, reachTask}
        \Else
            \State \Call{SignaturesExploration}{startEdge, $e$.edge, $e$.borders, spdi, reachTask}
        \EndIf
    \EndFor

    \For{$T$ \textbf{in} threads}
        \State \Call{JoinThread}{$T$}
        \State FreeThreads $\gets$ FreeThreads + 1
    \EndFor
\EndFunction
\end{algorithmic}
\end{algorithm}

\section{Benchmarks} \label{sec:benchmarks}

    We investigated several sources (\cite{speedi06}, \cite{CSMAFK15}, \cite{FI04}) where we looked for SPDI examples suitable to run benchmarks on. In the following list we present benchmarks we have managed to use in our experiments.

\begin{itemize}
    \item SPDI generated by MSPDI library \cite{speedi06}. This is a Perl library for generating SPDI files from 2-dimensional ordinary differential equations. Three examples we used include pendulum equations, spiral ODE with one focal point and the following non-linear system:
    \[
        \begin{cases}
            \dot{x} = y \\
            \dot{y} = -0.5 y - 2x - x^{2}
        \end{cases}
    \]
    \item The model example from \cite{speedi06}
    \item Randomly generated SPDIs (see \ref{random-spdi-generation})
\end{itemize}

\subsection{Random SPDI Generation} \label{random-spdi-generation}

    Here we introduce the algorithm for random SPDI generation. The motivation behind it is such that randomly generated examples are usually much more complex to solve and relatively easy to obtain. The algorithm is based on convex polygonal partitioning of a plane using Voronoi diagrams. It places $N$ random points on a $[0; 1000]^{2}$ square of $\mathbb{R}^{2}$ plane and creates a Voronoi diagram with this points as regions centers. The resulting partitioning is guaranteed to be convex based on the properties of Voronoi diagram (see Figure \ref{fig:spdi-example}, where SPDI is randomly generated).

    An edge of the region is called the \textit{output edge} if there is at least one trajectory that goes out of this region through this edge. Each edge in an SPDI could be an output edge for at most one region (we do not consider SPDIs with Zeno behaviour).
    
\begin{figure}[H]
        \centering
        \includegraphics[scale=0.60]{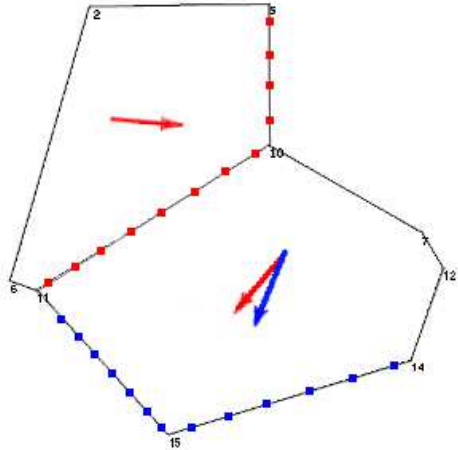}
        \captionsetup{width=.8\linewidth}
        \caption{Red dots denote the output edges for the upper region, and  blue dots denote the output edges for the lower region. The oriented angles between edges (6-11) and (2-5) for the upper region, and between edges (10-11) and (12-14) for the lower region, are positive.}
        \label{fig:output-edges}
    \end{figure}

    To define angles on a given partition, we assign a sequence of edges to be output edges for each region and test that the oriented angle between the pre-leftmost and the post-rightmost edges is positive (see Figure \ref{fig:output-edges}). If the output edges for all regions are correctly defined, we assign two vectors to each region by taking it randomly between the pre-leftmost and post-rightmost edge vectors.
    
    Output edges and vectors are assigned to each region by Algorithm \ref{alg:random-diff-inclusion}. Resulting SPDI is represented in the format proposed in \cite{speedi06}.

\begin{algorithm}
\caption{Constructing random differential inclusion}
\label{alg:random-diff-inclusion}
\begin{algorithmic}
\State Randomly iterate by all regions
\For{region $R$ \textbf{in} all regions, shuffled}
  \State find all starting edges $E$ \Comment {free edge with non-free previous neighbour}

  \If{$E$ is empty}
    \If{no free edges in $R$}
      \State \Return FAILURE
    \Else
      \State $E \gets$ \{random edge from $R$\} \Comment{all edges are free, no edge with non-free neighbour}
    \EndIf
  \EndIf

  \For{$e \in E$}
    \State try to construct an output set for $R$ starting with $e$

    \If{no output set is obtained}
      \State \Return FAILURE
    \Else
      \State assign two random vectors between pre-leftmost and post-rightmost edges to $R$
    \EndIf
  \EndFor
\EndFor
\end{algorithmic}
\end{algorithm}

    For benchmarking we implemented the random task generator which produces random reachability tasks for a given SPDI. It can also generate fixed sequences of random tasks, which is achieved by fixing a random seed. A reachability task is generated as a random set of start and final edge intervals. As the formats of reachability tasks in SPeeDI and ParaPlan differ, the generator produces the same set of tasks in both formats.

\section{Experiments}\label{sec:tests}

    We set up two series of experiments, one for comparing ParaPlan and SPeeDI, and the other for measuring the profit from paralellization. All our experiments were conducted on a 64-bit Linux computer with 8 core Intel Core-i7-4790T (2.7 GHz, 8MB cache) processor and 16 GB RAM. ParaPlan tool is implemented in C++ using Pthreads library for paralellization. Full code of the tool could be found online (see \cite{ST17}), as well as the SPDI benchmarks.

\subsection{Comparison of ParaPlan and SPeeDI}
 
    We compared ParaPlan and SPeeDI using the model SPDI example from \cite{speedi06}. For this purpose we ran a series of 100 and 1000 different reachability tasks. The results are presented in Table \ref{comparison-table}.

    \begin{table}[h]
    \caption{Comparison of ParaPlan and SPeeDI on 100 and 1000 tasks}\label{comparison-table} 
    \begin{center}
    \begin{tabular}{|l|l|l|}
    \hline
    Tool & 100 tasks & 1000 tasks \\ \hline
    ParaPlan & 0m 1.151s & 0m 9.804s \\ \hline
    SPeeDI & 1m 16.857s & 18m 40.828s \\ \hline
    \end{tabular}
    \end{center}
    \end{table}

    ParaPlan outperforms SPeeDI by 75-100 times on this example. Partly this result is achieved because our code is written in C++, while the original code is in Haskell, and partly due to the optimisation of the algorithm. We observed that in approximately 13\% cases the answers produced by the tools were different and we manually found several tasks on which SPeeDI produced an incorrect answer.

\subsection{Comparison of Parallel and Optimised Sequential Algorithms}

    Second experiment was aimed at revealing whether there is any profit that could be gained from parallelization. We generated a new series of reachability tasks for different SPDIs and compared the average time it took the ParaPlan tool to process all the tasks on different number of threads. The reachability tasks series for each SPDI was of length 100 and contained $1 \leqslant s \leqslant 10 $ starting edges and $1 \leqslant f \leqslant 10$ final edges, so that every combination of $s$ and $f$ is presented in tests.

    We divided all SPDIs onto two groups - the group of "heavy tests", where the average computational time on one thread was higher than 0.1 second, and the "light tests" group, where this time was less than 0.1 second. The group of heavy tests consists of SPDIs of various size, obtained from spiral ODE using MSPDI library. Light tests group was formed of example SPDI and randomly generated SPDI consisting of 100 regions. Other benchmarks were solved too fast to rely on the measured time and were not included in the final results table.

    Each group of tests was divided in two, depending on the reachability of the final set. It is done on purpose, because the algorithm terminates as soon as it reaches any point of final set, and in case of reachable tasks it happens much earlier than the whole DFS tree is looked up, which leads to much more considerable speed-up.

    We did not take into consideration those tests which did not finish in 5 seconds on at least one number of threads. In all other tests time is clipped in $[0, 5]$ seconds interval. Hence, the speed-up results we obtain is a lower bound for the real speed-up. We also did not perform an extreme test to determine how many regions in SPDI our tool can process, because it mainly depends on the complexity of the dynamics and not on the number of regions.

    In the following subsections we present the results of our testing. Each subsection contains two tables, one for the mean value of absolute time of testing, and the other for the relative speed-up observed. The data is presented also on two graphs in each subsection.

\subsubsection{Heavy Tests, Unreachable States}

    This is the hardest test for our tool because the whole DFS tree is needed to be looked up, and the SPDI is rather complex. However, we observe the growing speed-up of about 1.4 times at its peak.

\begin{table}[H]
\caption{Absolute testing time, mean value}
\begin{center}
\begin{tabular}{|l|l|l|l|l|l|l|l|l|}
\hline
Number of threads & 1 & 2 & 3 & 4 & 5 & 6 & 7 & 8 \\ \hline
spiral\_6  & 2.079 & 1.962 & 1.879 & 1.786 & 1.721 & 1.667 & 1.600 & 1.527 \\ \hline
spiral\_10 & 2.095 & 2.172 & 2.157 & 2.022 & 1.950 & 1.819 & 1.792 & 1.705 \\ \hline
spiral\_15 & 0.811 & 0.716 & 0.738 & 0.655 & 0.683 & 0.651 & 0.642 & 0.652 \\ \hline
\end{tabular}
\end{center}
\end{table}

\begin{table}[H]
\caption{Relative speed-up}
\begin{center}
\begin{tabular}{|l|l|l|l|l|l|l|l|l|}
\hline
Number of threads & 1 & 2 & 3 & 4 & 5 & 6 & 7 & 8 \\ \hline
spiral\_6  & 1.000 & 1.059 & 1.106 & 1.164 & 1.208 & 1.247 & 1.299 & 1.361 \\ \hline
spiral\_10 & 1.000 & 0.964 & 0.971 & 1.036 & 1.074 & 1.151 & 1.169 & 1.228 \\ \hline
spiral\_15 & 1.000 & 1.132 & 1.098 & 1.238 & 1.187 & 1.245 & 1.263 & 1.243 \\ \hline
\end{tabular}
\end{center}
\end{table}

\begin{figure}[H]
    \centering
    \begin{subfigure}{.5\textwidth}
        \centering
        \includegraphics[width=\linewidth]{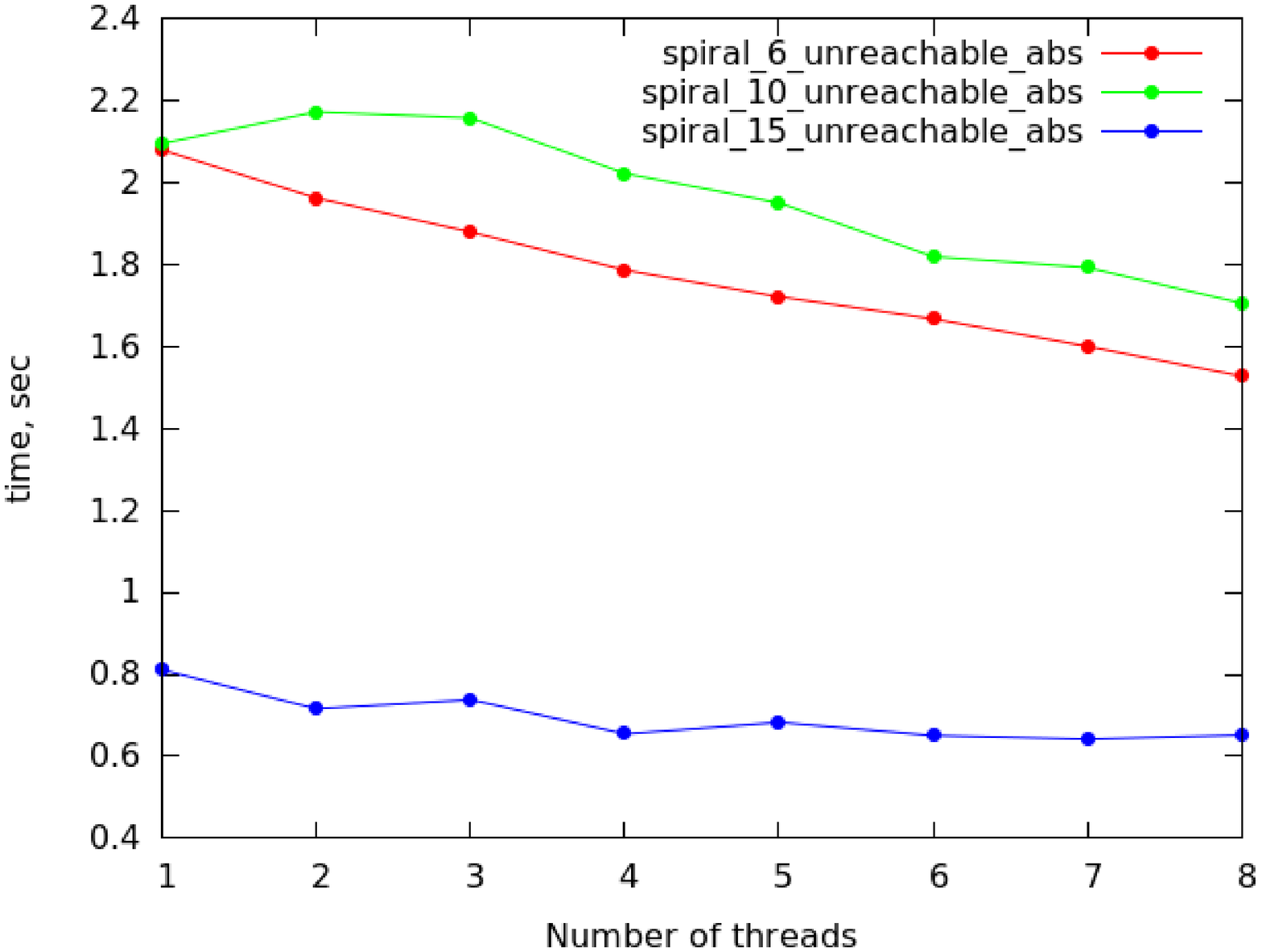}
        \captionsetup{width=.8\linewidth}
        \caption{Absolute testing time, mean value}
    \end{subfigure}%
    \begin{subfigure}{.5\textwidth}
        \centering
        \includegraphics[width=\linewidth]{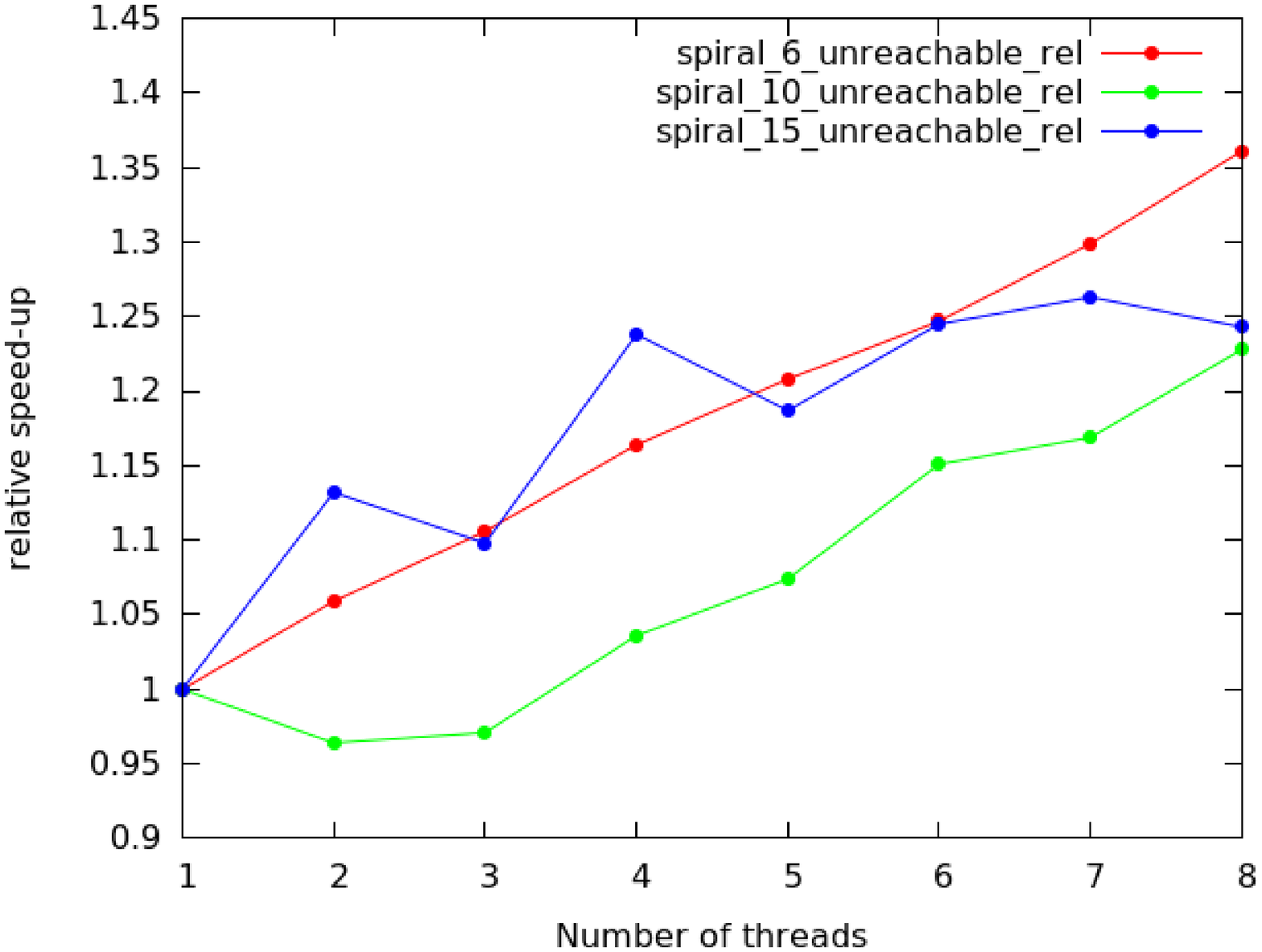}
        \captionsetup{width=.8\linewidth}
        \caption{Relative speed-up}
    \end{subfigure}%
\caption{Heavy tests, unreachable states}
\end{figure}

\subsubsection{Heavy Tests, Reachable States}

    Here we observe much more significant acceleration (up to 83 times faster), because often the algorithm finishes long before the whole DFS tree is done.

\begin{table}[H]
\caption{Absolute testing time, mean value}
\begin{center}
\begin{tabular}{|l|l|l|l|l|l|l|l|l|}
\hline
Number of threads & 1 & 2 & 3 & 4 & 5 & 6 & 7 & 8 \\ \hline
spiral\_6  & 0.909 & 0.735 & 0.442 & 0.276 & 0.163 & 0.105 & 0.074 & 0.068 \\ \hline
spiral\_10 & 3.378 & 2.033 & 1.232 & 0.599 & 0.333 & 0.171 & 0.110 & 0.074 \\ \hline
spiral\_15 & 4.245 & 2.431 & 1.376 & 0.828 & 0.417 & 0.195 & 0.159 & 0.051 \\ \hline
\end{tabular}
\end{center}
\end{table}

\begin{table}[H]
\caption{Relative speed-up}
\begin{center}
\begin{tabular}{|l|l|l|l|l|l|l|l|l|}
\hline
Number of threads & 1 & 2 & 3 & 4 & 5 & 6 & 7 & 8 \\ \hline
spiral\_6  & 1.000 & 1.236 & 2.056 & 3.293 & 5.576 & 8.657 & 12.28 & 13.36 \\ \hline
spiral\_10 & 1.000 & 1.661 & 2.741 & 5.639 & 10.14 & 19.75 & 30.70 & 45.64 \\ \hline
spiral\_15 & 1.000 & 1.746 & 3.085 & 5.126 & 10.17 & 21.76 & 26.69 & 83.23 \\ \hline
\end{tabular}
\end{center}
\end{table}

\begin{figure}[H]
    \centering
    \begin{subfigure}{.5\textwidth}
        \centering
        \includegraphics[width=\linewidth]{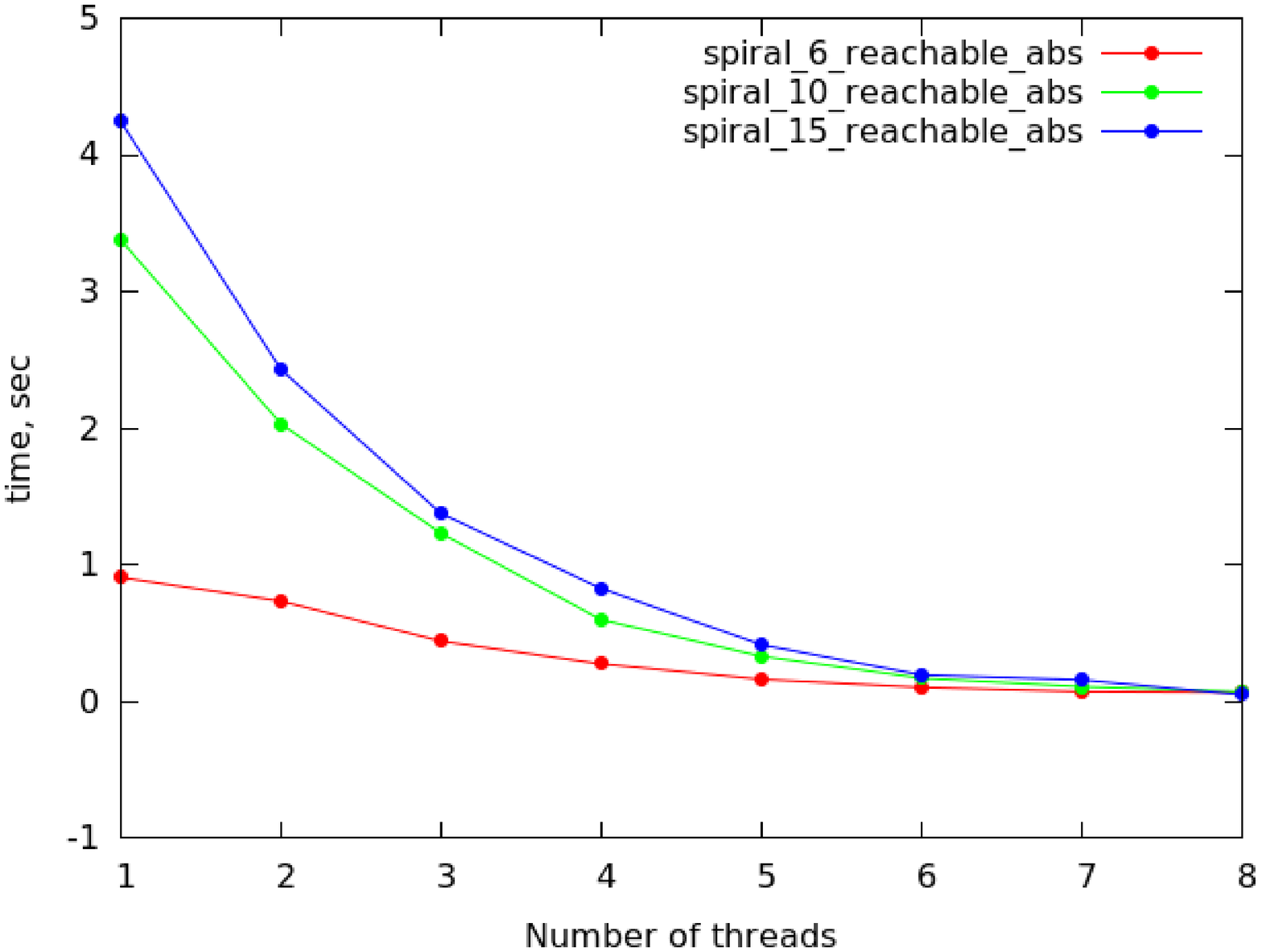}
        \captionsetup{width=.8\linewidth}
        \caption{Absolute testing time, mean value}
    \end{subfigure}%
    \begin{subfigure}{.5\textwidth}
        \centering
        \includegraphics[width=\linewidth]{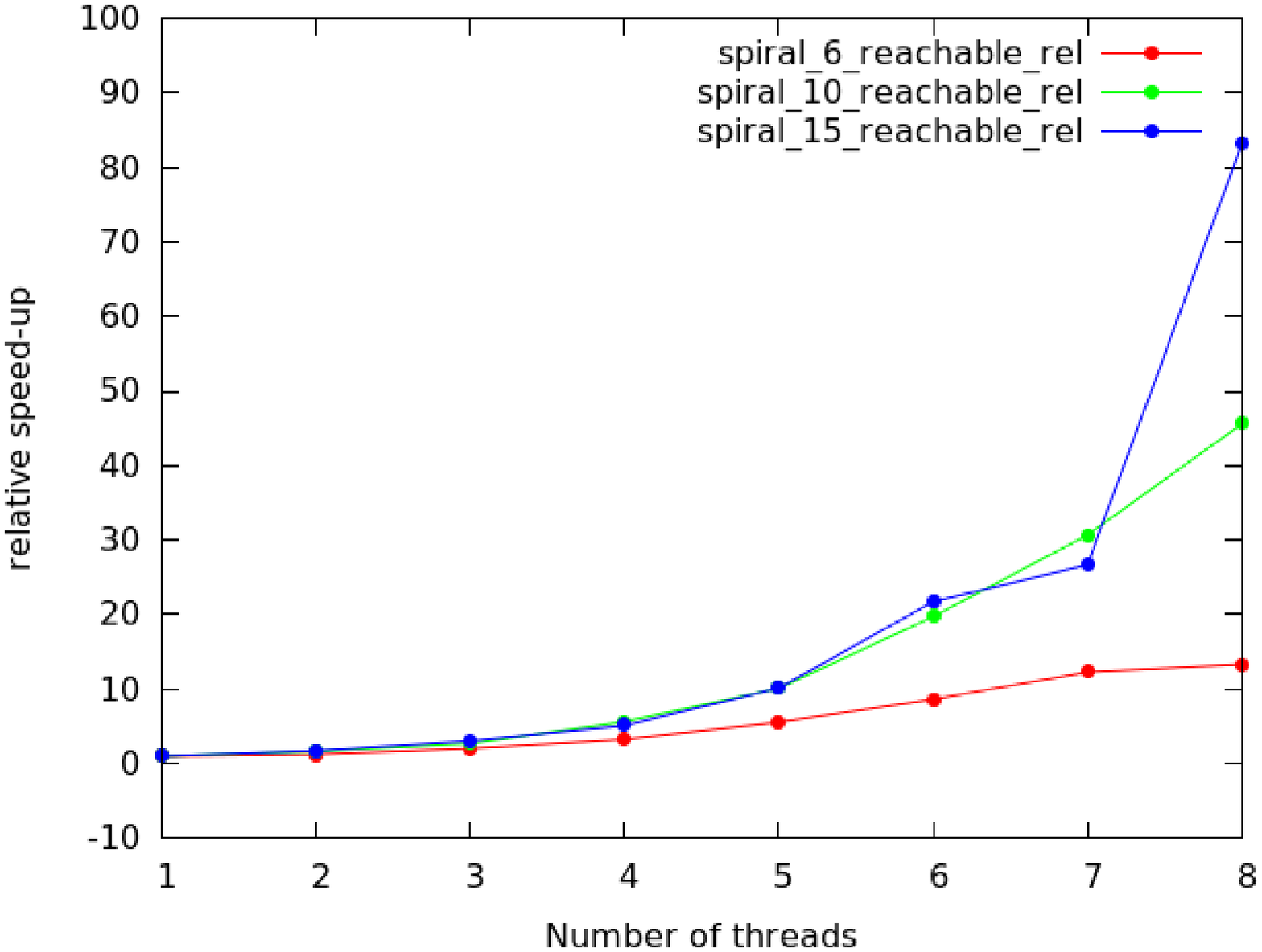}
        \captionsetup{width=.8\linewidth}
        \caption{Relative speed-up}
    \end{subfigure}%
\caption{Heavy tests, reachable states}
\end{figure}

\subsubsection{Light Tests, Unreachable States}

    On light tests there is practically no effect of paralellization, as the tasks finish very fast. There is even a little slowdown on random SPDI because of the threads overhead. We also observe that against our expectations random SPDIs happened to be relatively easy to solve for our tool.

\begin{table}[H]
\caption{Absolute testing time, mean value}
\begin{center}
\begin{tabular}{|l|l|l|l|l|l|l|l|l|}
\hline
Number of threads & 1 & 2 & 3 & 4 & 5 & 6 & 7 & 8 \\ \hline
example & 0.020 & 0.018 & 0.017 & 0.018 & 0.018 & 0.018 & 0.019 & 0.018 \\ \hline
random\_100 & 0.014 & 0.013 & 0.014 & 0.015 & 0.015 & 0.015 & 0.015 & 0.015 \\ \hline
\end{tabular}
\end{center}
\end{table}

\begin{table}[H]
\caption{Relative speed-up}
\begin{center}
\begin{tabular}{|l|l|l|l|l|l|l|l|l|}
\hline
Number of threads & 1 & 2 & 3 & 4 & 5 & 6 & 7 & 8 \\ \hline
example & 1.0 & 1.111 & 1.176 & 1.111 & 1.111 & 1.111 & 1.052 & 1.111 \\ \hline
random\_100 & 1.0 & 1.076 & 1.0 & 0.933 & 0.933 & 0.933 & 0.933 & 0.933 \\ \hline
\end{tabular}
\end{center}
\end{table}

\begin{figure}[H]
    \centering
    \begin{subfigure}{.5\textwidth}
        \centering
        \includegraphics[width=\linewidth]{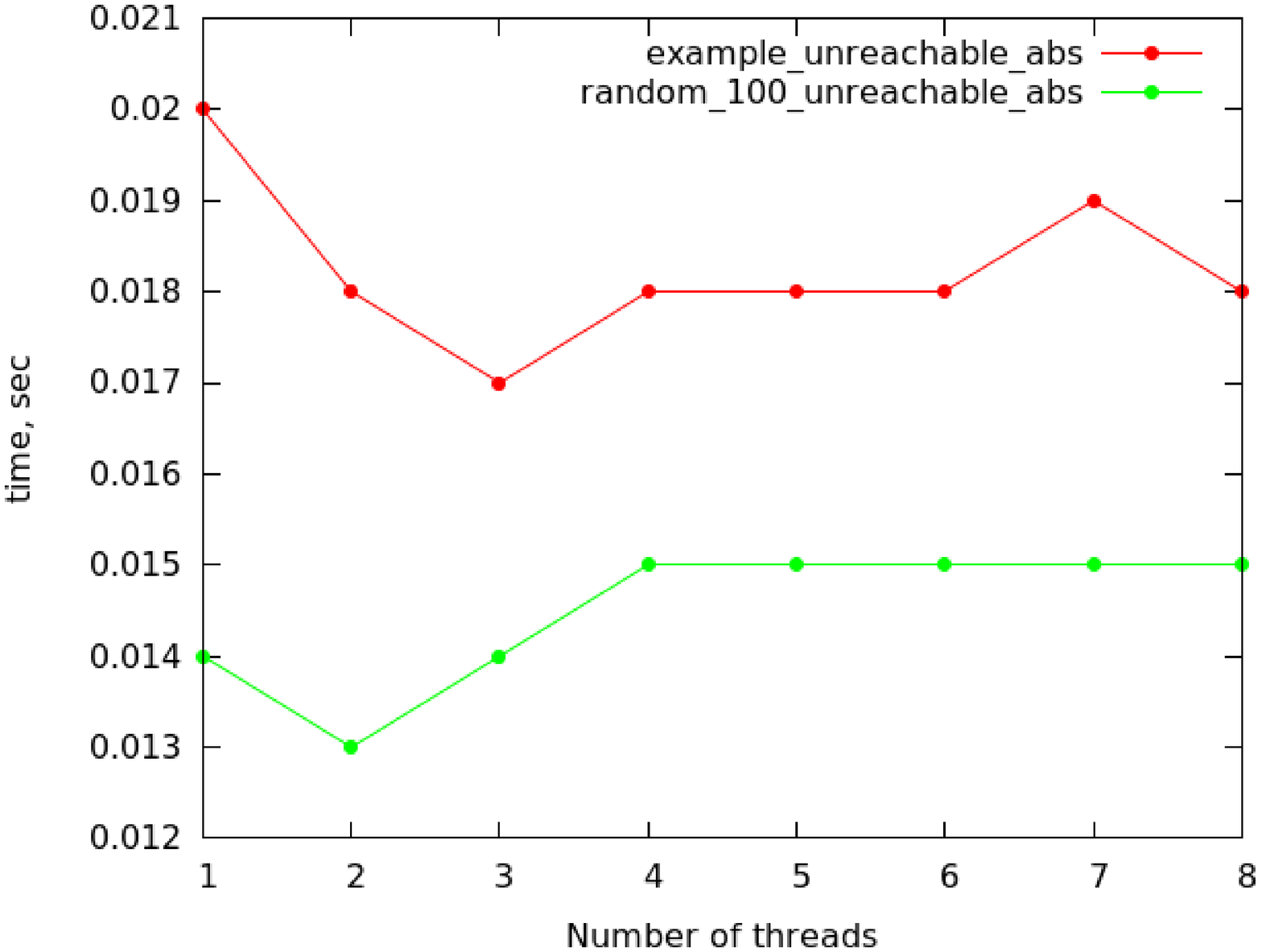}
        \captionsetup{width=.8\linewidth}
        \caption{Absolute testing time, mean value}
    \end{subfigure}%
    \begin{subfigure}{.5\textwidth}
        \centering
        \includegraphics[width=\linewidth]{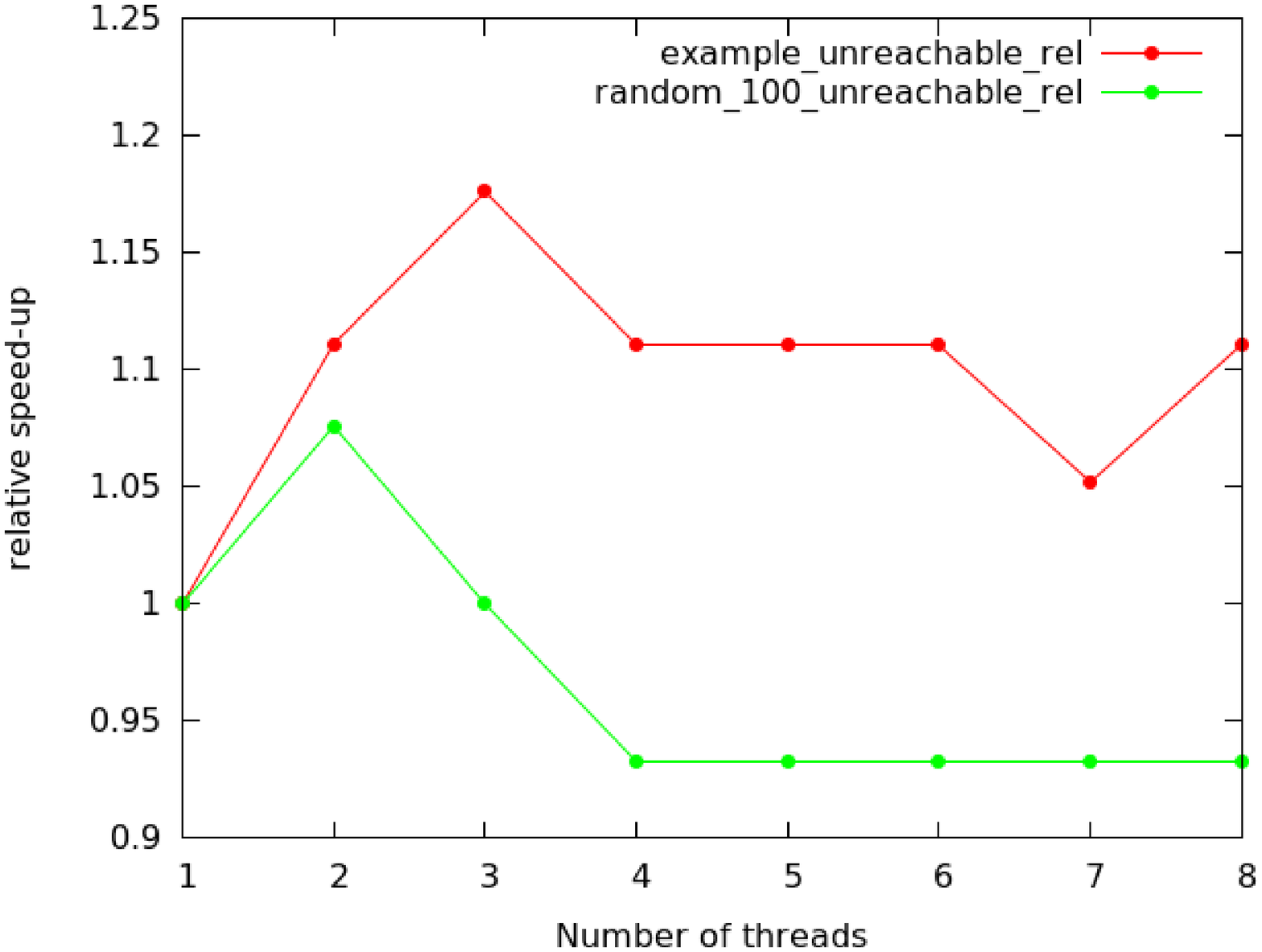}
        \captionsetup{width=.8\linewidth}       \caption{Relative speed-up}
    \end{subfigure}%
\caption{Light tests, unreachable states}
\end{figure}

\subsubsection{Light Tests, Reachable States}

    Here we see little to none effect of parallelization, and again a little slowdown on random SPDI, and a slight speed-up on the model example, which stops on the 6 threads.

\begin{table}[H]
\caption{Absolute testing time, mean value}
\begin{center}
\begin{tabular}{|l|l|l|l|l|l|l|l|l|}
\hline
Number of threads & 1 & 2 & 3 & 4 & 5 & 6 & 7 & 8 \\ \hline
example & 0.016 & 0.011 & 0.008 & 0.009 & 0.009 & 0.007 & 0.007 & 0.007 \\ \hline
random\_100 & 0.013 & 0.013 & 0.013 & 0.014 & 0.015 & 0.014 & 0.014 & 0.014 \\ \hline
\end{tabular}
\end{center}
\end{table}

\begin{table}[H]
\caption{Relative speed-up}
\begin{center}
\begin{tabular}{|l|l|l|l|l|l|l|l|l|}
\hline
Number of threads & 1 & 2 & 3 & 4 & 5 & 6 & 7 & 8 \\ \hline
example & 1.0 & 1.454 & 2.0 & 1.777 & 1.777 & 2.285 & 2.285 & 2.285 \\ \hline
random\_100 & 1.0 & 1.0 & 1.0 & 0.928 & 0.866 & 0.928 & 0.928 & 0.928 \\ \hline
\end{tabular}
\end{center}
\end{table}

\vspace{-0.3cm}

\begin{figure}[H]
    \centering
    \begin{subfigure}{.5\textwidth}
        \centering
        \includegraphics[width=\linewidth]{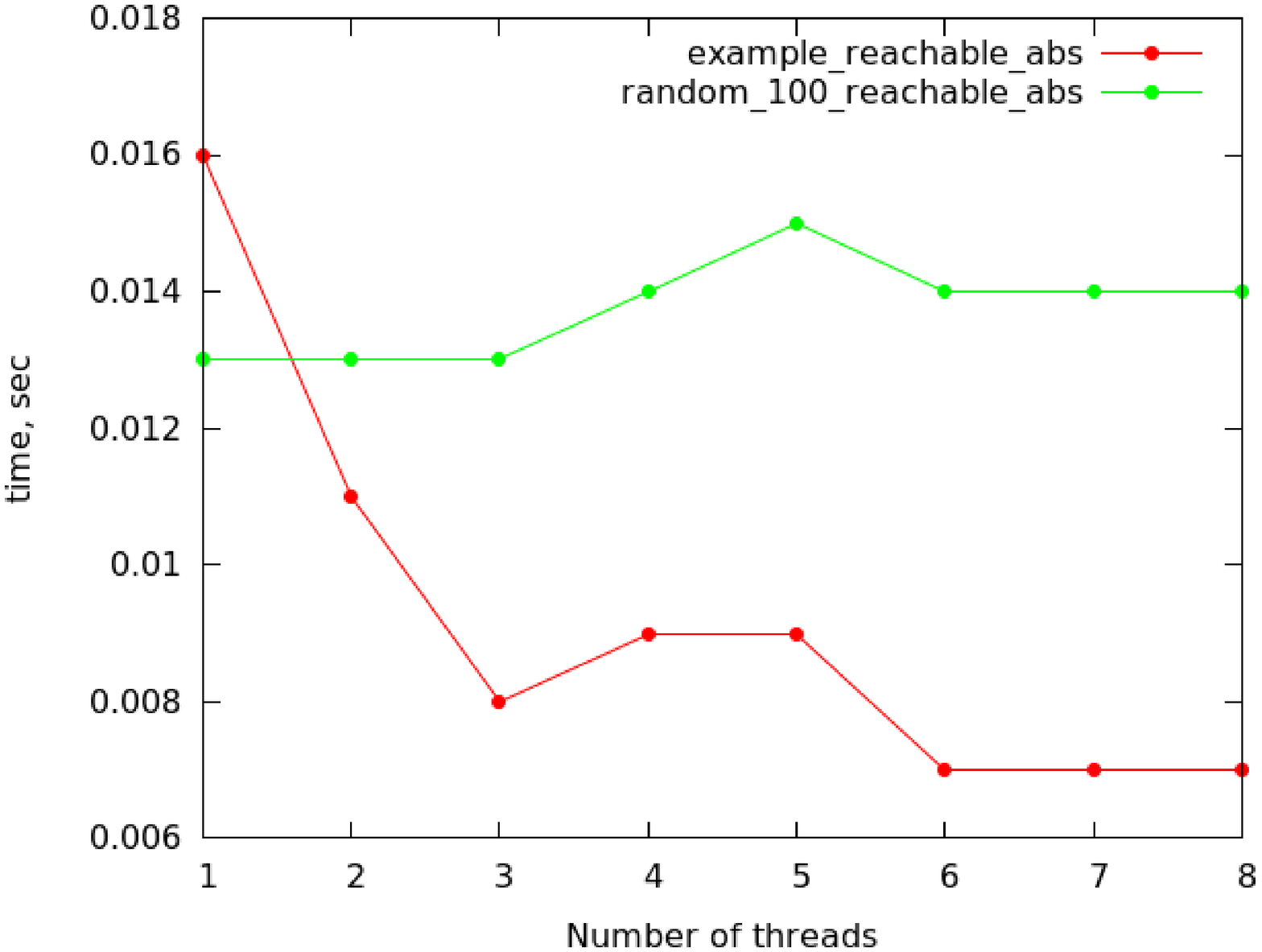}
        \captionsetup{width=.8\linewidth}
        \caption{Absolute testing time, mean value}
    \end{subfigure}%
    \begin{subfigure}{.5\textwidth}
        \centering
        \includegraphics[width=\linewidth]{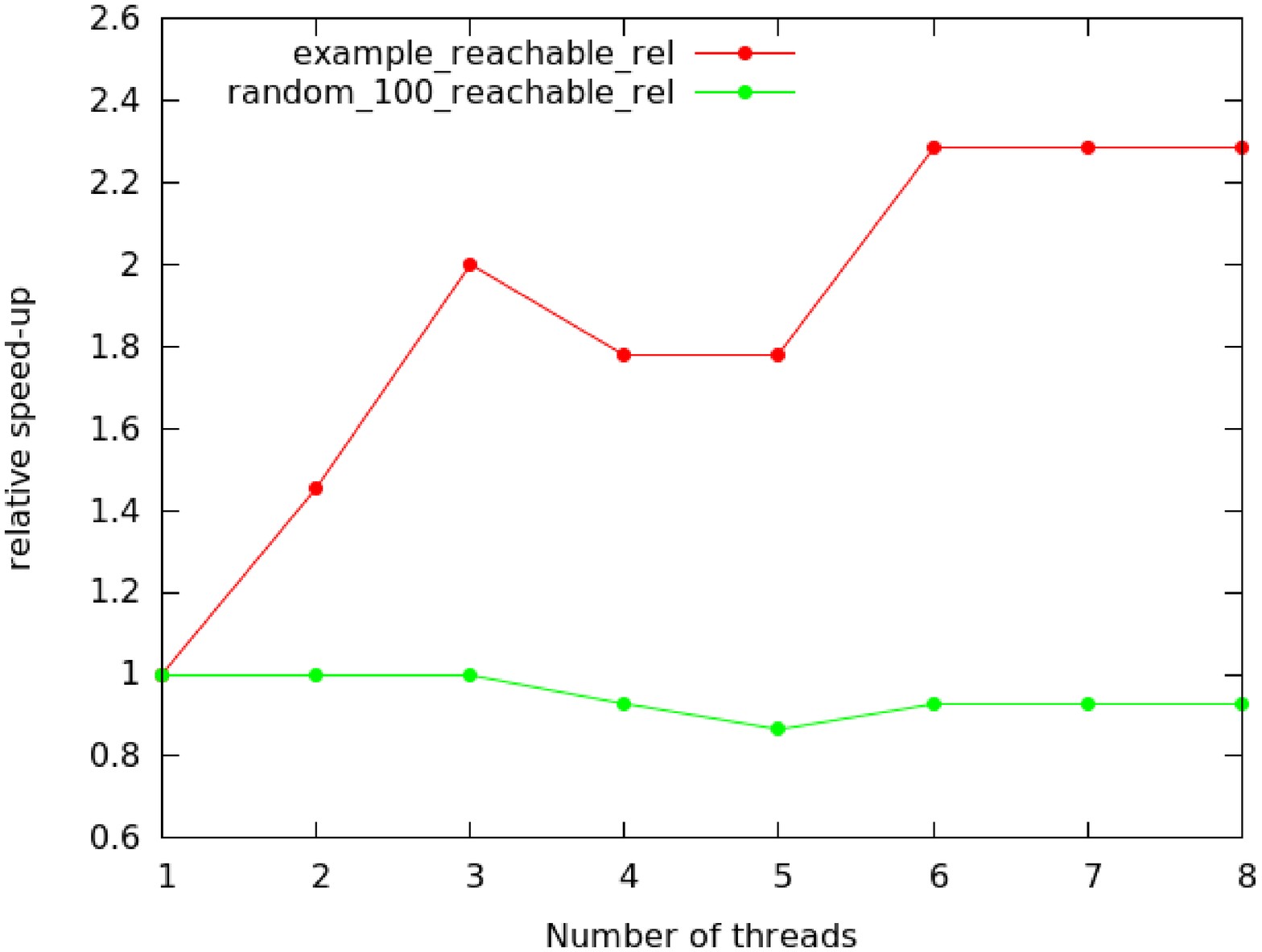}
        \captionsetup{width=.8\linewidth}
        \caption{Relative speed-up}
    \end{subfigure}%
\caption{Light tests, reachable states}
\end{figure}

\section{Conclusion and Future Work}\label{sec:conclusions}

    We presented the paralellized algorithm for reachability analysis of 2-dimensional systems of polygonal differential inclusions. The algorithm is the optimized version of the original algorithm from \cite{ASY07}. It was experimentally demonstrated that a speed-up could be gained via parallelization, which depends on the complexity of an SPDI itself and on whether or not the final states are reachable. We also presented the algorithm for generating random SPDIs which could be useful for future research in this area.
Possible future work may be focused on representing SPDIs as general hybrid automata and comparing the results of solving reachability tasks on SPDIs using ParaPlan and the existing model checkers for hybrid systems.

\bibliographystyle{eptcs}
\bibliography{references}
\end{document}